\documentclass[twocolumn]{IEEEtran}
\usepackage{cite}
\usepackage{setspace}
\usepackage{mathtools}
\usepackage{graphicx}
\usepackage{caption}
\usepackage{subcaption}
\usepackage{float}
\floatstyle{plaintop}
\restylefloat{table}
\usepackage[justification=centering]{caption}
\usepackage{filecontents}
\usepackage{tablefootnote}
\usepackage{fnpos}
\usepackage{booktabs}
\usepackage{mathtools}

\begin{document}
\title{ Energy Efficiency in Massive MIMO-Based 5G Networks: Opportunities and Challenges}

\author{K.~N.~R.~Surya~Vara~Prasad, Ekram~Hossain, and Vijay~K.~Bhargava}
\vspace{-5mm}
\maketitle
\thanks{}

\begin{abstract}
As we make progress towards the era of fifth generation (5G) communication networks, energy efficiency (EE) becomes an important design criterion because it guarantees sustainable evolution. In this regard, the massive multiple-input multiple-output (MIMO) technology, where the base stations (BSs) are equipped with a large number of antennas so as to achieve multiple orders of spectral and energy efficiency gains, will be a key technology enabler for 5G. In this article, we present a comprehensive discussion on state-of-the-art techniques which further enhance the EE gains offered by massive MIMO (MM). We begin with an overview of MM systems and discuss how realistic power consumption models can be developed for these systems. Thereby, we discuss and identify few shortcomings of some of the most prominent EE-maximization techniques present in the current literature. Then, we  discuss ``hybrid MM systems" operating in a 5G architecture, where MM operates in conjunction with other potential technology enablers, such as millimetre wave, heterogenous networks, and energy harvesting networks. Multiple opportunities and challenges arise in such a 5G architecture because these technologies benefit mutually from each other and their coexistence introduces several new constraints on the design of 
energy-efficient systems. Despite clear evidence that hybrid MM systems can achieve significantly higher EE gains than conventional MM systems, several open research problems continue to roadblock system designers from fully harnessing the EE gains offered by hybrid MM systems. Our discussions lead to the conclusion that hybrid MM systems offer a sustainable evolution towards 5G networks and are therefore an important research topic for future work.

\end{abstract}


\section{Introduction}
\subsection {Expectations from 5G Cellular Networks}
The information and communication technology (ICT) industry is making rapid progress toward the fifth generation (5G) networks, which are expected to integrate almost anything and everything across the globe into the Internet. A host of inter-connected networks including smart cities, vehicular networks, and augmented reality hubs will co-exist within 5G. In terms of technology demands, 5G networks should provide peak data rates up to 20 Gbps,  average data rates greater than 100 Mbps, and seamless connectivity for a huge number of Internet-of-Things (IoT) devices per square kilometre.

While planning for 5G networks, energy consumption becomes a critical concern because mobile communication networks contribute significantly towards the global carbon footprint. Trends \cite{MM1} suggest that the ICT sector would emit more than 250 million tonnes of greenhouse gas per annum by 2020. Therefore, to ensure sustainability, 5G networks should operate at low energy consumption levels while still achieving large capacity gains. An important design criterion in this context is the bit-per-joule energy efficiency (EE), defined as

\begin{equation} \label{eq1}
EE = \frac {R} {P}, 
\end {equation}

\noindent where $R$ is the system throughput and $P$ is the power spent in achieving $R$. The recently proposed massive multiple-input multiple-output (MIMO) technology offers multiple orders of spectral and energy efficiency gains over current LTE technologies, and is therefore, a promising enabler for 5G.

\begin{figure}
\begin{center}
\includegraphics[scale=0.5]{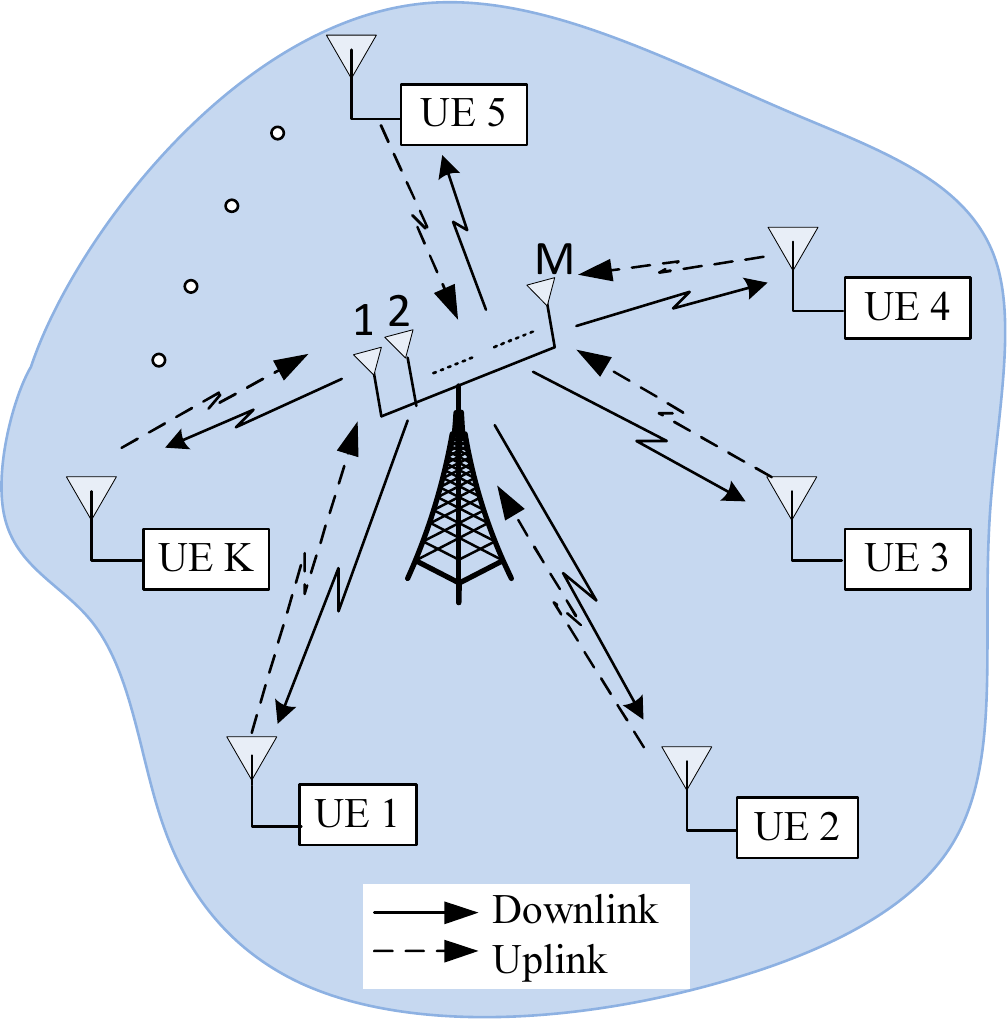}
\caption{Massive MIMO: A multi-user MIMO technology where $K$ single-antenna UEs are served by a BS with $M >> K$ antennas.}
\label{fig_mu_mimo}
\end{center}
\end{figure}

\subsection {Overview of Massive MIMO Technology} \label{sec_mm}

Massive MIMO (MM) is a multi-user MIMO technology in which $K$ single-antenna user equipments (UEs) are serviced on the same time-frequency resource by a base station (BS) equipped with a relatively large number $M$ of antennas, i.e., $M >> K$ (c.f. Fig. \ref{fig_mu_mimo}).

Deploying several antennas at the BS results in an interesting propagation scenario called {\em favourable propagation}, where the channel becomes near-deterministic because the BS-UE radio links become nearly orthogonal to each other \cite{MM2}. This is  because the effects of fast fading, intra cell interference, and uncorrelated noise disappear asymptotically in the large $M$ regime. Significant EE gains can be achieved under {\em favourable propagation} because multiple orders of multiplexing and array gains are realizable. For illustration, let us discuss the uplink (UL) and downlink (DL) transmissions in an MM cell.

The asymptotic Shannon capacities on the UL ($C_{UL}$) and the DL ($C_{DL}$) for an MU-MIMO channel under {\em favourable propagation} are given by \cite{newMM3}

\begin {equation} \label {shannon_cap_eq}
\begin {aligned}
C_{UL} = \sum_{k = 1}^{K} \log_2 (1 + p_u M \beta_k), \\
C_{DL} = \underset{(a_k \geq 0, \sum a_k \leq 1)}{\max} \sum_{k = 1}^{K} \log_2 (1 + p_d M a_k \beta_k),
\end{aligned}
\end {equation}

\noindent where $p_u$ and $p_d$ are the average UL and DL signal to noise ratios (SNRs), $\{\beta_k\}, k = 1, \cdots, K$, represents the large-scale fading coefficients for the $K$ UEs, and $\{a_k\}$ is a set of variables which should be optimized to obtain $C_{DL}$. When appropriate power control strategies are used to normalize the effect of $\beta_k$, the UL capacity simplifies to $K \log_2 (1 + M p_u)$. A similar observation can be made on the DL. This clearly shows that we can achieve $O(K)$ multiplexing gains and $O(M)$ array gains under {\em favourable propagation}. In fact, these gains can be achieved using simple linear processing techniques at the BS, such as, maximal ratio combining (MRC)  and zero forcing (ZF) detection. This greatly simplifies the computational load and the hardware requirements at the BSs because conventionally, the BSs implement complex signal processing techniques, such as maximum likelihood (ML) detection and successive interference cancellation (SIC), to achieve optimal capacities. Since the computational requirements are relaxed considerably, we observe a drastic reduction in the circuit power consumed by the system. MM delivers multiple orders of EE gains because it offers large multiplexing and array gains at reduced power consumption levels.

\subsection {How Practical is Massive MIMO?}

{\em Favorable propagation} is derived in \cite {MM2} as an asymptotic propagation scenario for independent and identically distributed (i.i.d) Rayleigh channels, achieved when $M$ is increased unboundedly. There is a general notion that massive MIMO may not be practical because some of the assumptions behind {\em favourable propagation} may not be valid in practice. For example, it may not be feasible to increase $M$ unboundedly. Also, due to rich scattering environments, practical channels are known to exhibit fundamentally different propagation characteristics when compared to theoretical i.i.d Rayleigh channels. Despite these concerns, recent field studies \cite{MM3} show that measured channels with large, but finite, $M$, achieve a significant portion of the multiplexing and array gains derived under theoretical assumptions. This shows that massive MIMO is indeed a practical technology. 

Nevertheless, several technological challenges continue to exist. For example, designing compact MM antenna arrays is a challenge at the current sub-3GHz bands because a minimum inter-antenna spacing of $\lambda/2$, where $\lambda$ is the carrier wavelength, is required to avoid spatial correlation. Also, pilot-aided channel estimation techniques, which are widely used in the current LTE networks, cannot be applied to MM systems because the training overhead grows linearly with $M$ and $K$. Pilot contamination, which arises when non-orthogonal pilots are used for UL channel estimation, continues to be a major limiting factor on the performance of MM systems.

\section {Power Consumption in Massive MIMO Systems} \label {EE_define}

The sum power consumption $P$, aggregated over UL and DL transmissions in an MM system, can be modelled as
\begin{equation} \label{eq3}
P = P_{PA} +P_{C} + P_{sys},
\end {equation}

\noindent where $P_{PA}$ represents the total UL and DL power consumed by the power amplifiers (PAs) at the BS and the UEs, $P_{C}$ represents the total UL and DL circuit power consumed by different analog and digital signal processing circuits at the BS and the UEs, and $P_{sys}$ refers to the remaining system dependant component in $P$. While $P_{PA}$ accounts for the sum power expenditure on RF transmissions, $P_C$ includes the sum power consumption from RF chain components, such as, filters, mixers, and synthesizers, as well as baseband operations, such as, digital up/down conversion, FFT/IFFT, precoding/receiver combining, channel coding/decoding, and channel estimation. Note that $P_C$ cannot be modelled as per conventional practice as a constant term independent of ($M, K$) because the hardware requirements and the number of circuit operations in the system grow with $M$ and $K$. For example, with the {\em one RF chain per antenna} design used in the current LTE networks, the number of RF chains at the BS and the UEs grows affinely with $M$ and $K$, respectively. Additionally, the computational requirements for various baseband operations are functions in $M$ and $K$. For example \cite{MM20}, $O(M, K^2)$ operations are required for ZF precoding, $O(MK)$ for MRC detection, $O(MK)$ for minimum mean squared error (MMSE) channel estimation, and $O(K)$ for channel coding respectively. Therefore, realistic models should treat $P_C$ as a function in ($M,K$) and the variability of $P_C$ with ($M$, $K$) should be investigated during the design of energy-efficient MM networks. Lastly, $P_{sys}$ accounts for the power consumed by site-specific and architecture-specific factors, such as, BS and UE architectures, power supply, cooling system, backhaul, and other control equipment. $P_{sys}$ will play a significant role in characterizing EE for 5G networks because several BS and UE types will co-exist in a multi-tier architecture with different cell sizes, power consumption levels, and access technologies. 

\section {Designing Energy-Efficient Massive MIMO Systems} \label{ee_design_methods}

\par As we can observe from (\ref{eq1}), the energy efficiency of an MM system can be maximized by achieving near-optimal throughput performance at reduced power consumption levels. Based on this analogy, a number of research directions have been pursued for the design of energy-efficient MM networks. Few methods devise low complexity algorithms for BS operations such as multi-user detection, precoding, and user scheduling, so as to minimize power expenditure in the system. Few other methods, such as, transceiver redesign, antenna selection, and power amplifier dimensioning, focus on improving resource utilization so as to relax hardware requirements and hence, power expenditure in the system. Available literature also includes methods which introduce hardware imperfections so as to reduce power expenditure in the system. This section explores some of the most prominent EE-maximization techniques for MM networks and identifies few open research challenges thereof.

\subsection {Low-complexity BS operations} \label{sec_fdd}
Due to {\em favourable propagation} in the large $M$ regime, simple linear processing techniques, such as MRC detection and Maximal Ratio Transmission (MRT) precoding, and simple user scheduling algorithms, such as random and round robin scheduling, achieve near-optimal throughput performance. These simplifications yield significant EE gains because the circuit power $P_C$ is drastically reduced when compared to conventional systems with computationally intensive signal processing schemes, such as ML detection and SIC, and complex scheduling algorithms, such as random beamforming and semi-orthogonal user selection. 

While channel reciprocity can be exploited in TDD systems to derive near-optimal low-complexity linear precoding schemes, precoders for FDD systems cannot exploit channel reciprocity because the UL and DL communications occur on separate frequency bands. FDD precoders cannot also rely on pilot signalling and feedback from the UEs because this consumes at least $M + K$ symbols per coherence interval, making them impractical for high mobility scenarios. Few low overhead FDD precoders, which assume channel sparsity for channel dimensionality reduction, have been proposed  recently\cite{MM4}. However, such precoders are limited to high frequency bands, such as mmWave, where channel sparsity assumptions are valid. Consequently, low-complexity FDD precoding continues to be a major research challenge for MM networks. Since there are many more licenses worldwide for FDD than for TDD, progress on low overhead FDD precoders will promote wider acceptance of MM as a future technology.

\begin{figure}
\begin{center}
\includegraphics [scale=0.5] {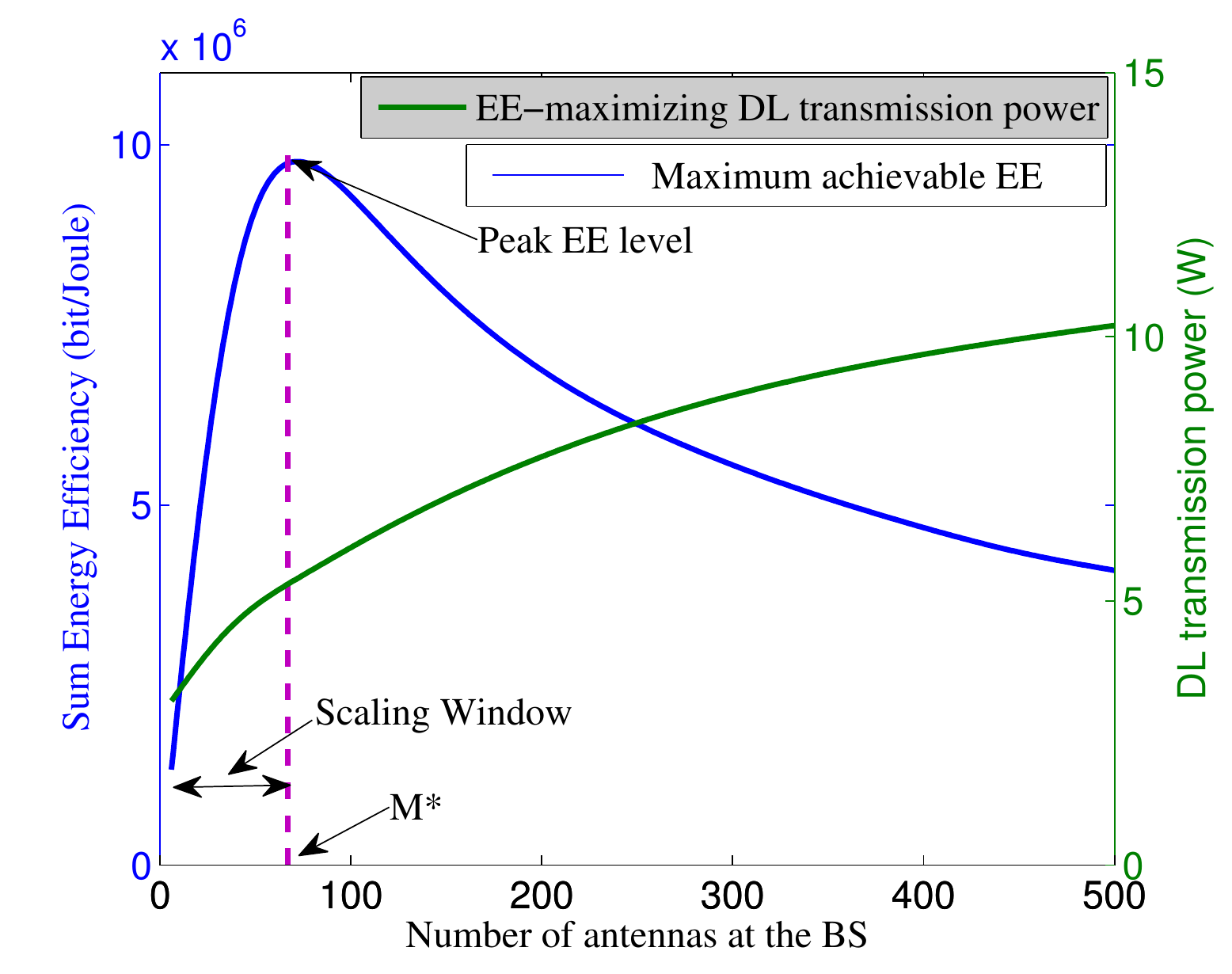}
\caption{Scaling the number antennas at the BS to improve energy-efficiency in massive MIMO systems.}
\label{fig_m_scaling}
\end{center}
\end{figure}

\subsection {Scale the Number of BS Antennas}
When the number of antennas at the BS is increased, the system throughput $R$ can be improved because higher multiplexing gains are achievable. However, observe from our discussions in the previous section that the circuit power $P_C$ also increases with M. Nevertheless, increasing $M$ can still be an energy-efficient strategy for MM networks if we somehow increase $R$ sufficiently that it dominates the increase in $P_C$. Although not obvious from initial observations, this can be done by increasing the DL transmission power over a certain scaling window. To understand why, let us study how DL transmission power can be optimized for energy efficiency. 

As we can observe from (\ref{eq1})-(\ref{eq3}), the system EE is a non-linear function in the DL transmission power because the EE metric takes a fractional form, where the numerator, i.e, the system throughput $R$, and the denominator, i.e,  the power consumption $P$, are both functions in the DL transmission power. Therefore, to optimize the DL transmission power for energy efficiency, we can use standard non-linear optimization methods, such as gradient descent \cite{MM20}. Optimal DL transmission powers would be different for different $(M, K)$ because the system EE depends on $(M, K)$ as well (c.f. Section \ref{sec_mm}, \ref{EE_define}). Using this methodology, we plot in Fig. \ref{fig_m_scaling}, the maximum achievable EE values and the corresponding DL transmission powers when $M$ is increased from $5$ to $500$. The simulation assumes $K = 30$, MRC detection on the UL, maximal ratio transmission (MRT) precoding on the DL, and a power consumption model based on prior works \cite{MM5}.

Clearly, we observe that there exists a scaling window, within which the system EE can be increased by simultaneously increasing $M$ and the DL transmission power. The scaling window is governed by a certain threshold $M^*$ on the number of BS antennas, beyond which $R$ approaches near-optimal performance bounds but $P_C$ continues to grow unboundedly with $M$. As a result, we observe that the system EE attains a peak level at $M^*$ and decreased gradually with $M$, even if we increase the DL transmission power. Note that the scaling window can be expanded by reducing the RF chain requirements at the BS because this results in reduced $P_C$ levels. Alternatively, if the transceiver design ensures very low circuit power expenditure, i.e., if $P_C << P$, it may be possible to reduce the DL transmission power with $M$ and still achieve improvements in the system EE. However, such a scenario relies heavily on the future of energy-aware transceiver design.

\subsection {Minimize PA Power Losses}
Significant EE gains can also be achieved by minimizing PA power losses because inefficient PA operations discard as much as 60\% to 95\% of the power input to the PAs. PA power losses can be minimized by operating the PAs at points close to the maximum allowed output. Unfortunately, most PAs in current deployments are operated at points much lower than the maximum allowed output because of the high linearity requirements imposed by high peak to average power (PAPR) waveforms such as OFDM. While few low PAPR non-orthogonal waveforms, such as single carrier modulation (SCM), have been proposed recently, designing appropriate non-orthogonal waveforms continues to be a major research challenge because most of the recently proposed waveforms suffer from limitations, such as long filter lengths and complex receiver techniques. Alternatively, PA linearity requirements can also be relaxed using constant envelope signals but generating such signals is an unresolved challenge till date.

\subsection {Minimize RF Chain Requirements at the BS} \label{sec_rf_chain}
Conventionally, MIMO precoding and beamforming are performed digitally in the baseband. Since digital processing requires dedicated baseband and RF chain components for each antenna element, BS transceivers conventionally adopt a {\em one RF chain per antenna} design. However, such a design results in significant circuit power consumption in the MM regime because the number of RF chains at the BS increases affinely with $M$. Therefore, minimizing RF chain requirements at the BS is an attractive strategy to improve EE in MM networks. Prominent techniques which reduce RF chain requirements include hybrid precoding, antenna selection, and transceiver redesign. Hybrid precoding techniques are generally built on channel sparsity assumptions and are hence, discussed in the context of mmWave systems in Section \ref{sec_mmwave}.

\subsubsection {Antenna selection}
\par Antenna selection is a signal processing technique which improves throughput in a system while simultaneously reducing the number of RF chains at the BS. Basically, a subset comprising $N$ out of the $M$ BS antennas is selected based on a predefined selection criterion, for example, to maximize throughput, SNR, or EE. Antennas in the selected subset are then connected to RF chains for further processing. Since the number of RF chains is reduced from $M$ to $N$, circuit power consumption in the system is reduced. 

\begin{figure}
\begin{center}
\includegraphics [scale=0.5] {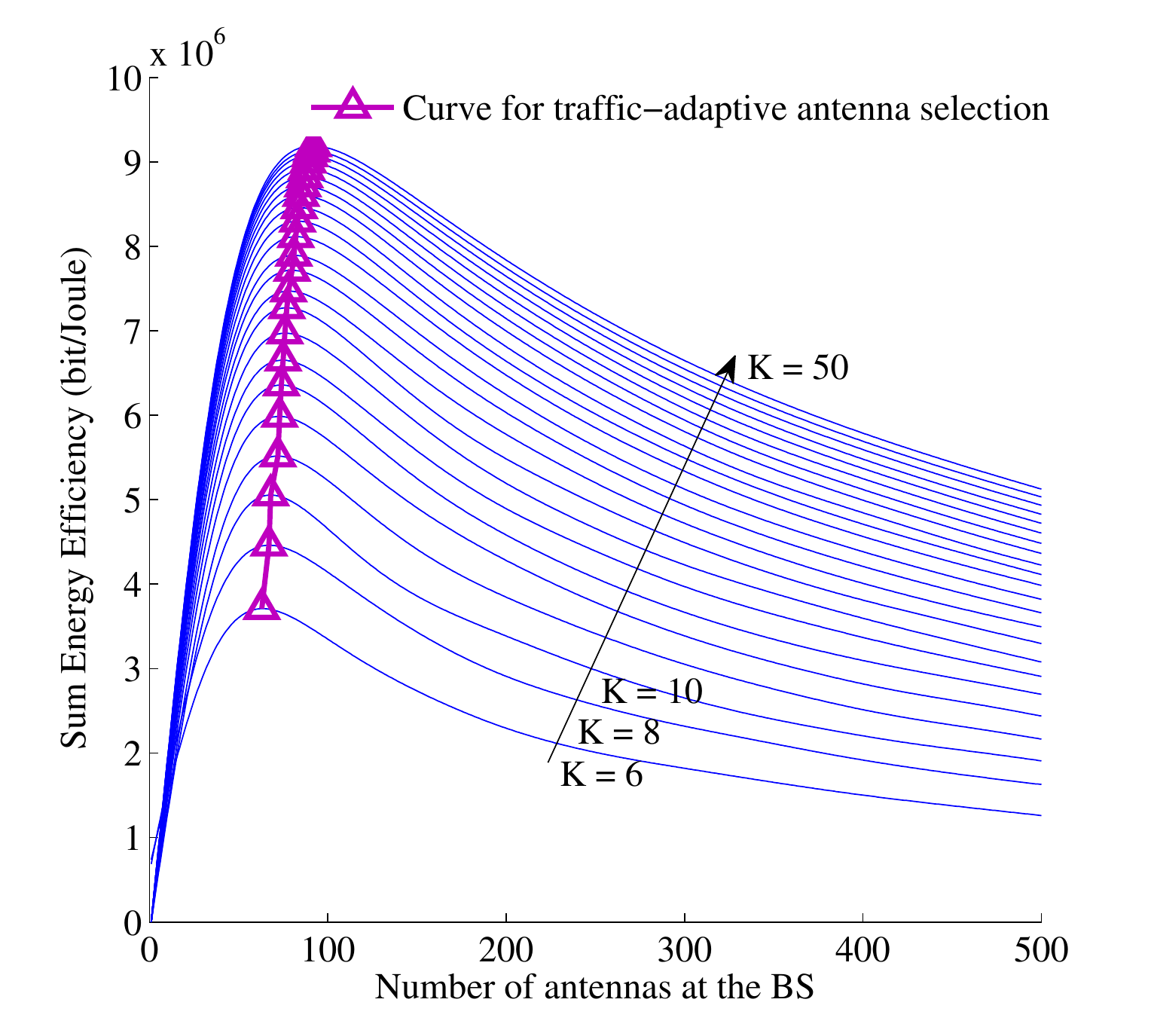}
\caption{Traffic-adaptive antenna selection for energy efficiency in massive MIMO systems.}
\label{fig_antenna_selection}
\end{center}
\end{figure}

Fig. \ref{fig_antenna_selection} provides a guideline to design traffic-adaptive antenna selection methods for energy efficiency in a massive MIMO system. To study how the system EE varies with traffic demands, we plot the maximum achievable EE as in Fig. \ref{fig_m_scaling}, but for $K$ values ranging from $6$ to $50$. To obtain the maximum EE values for each $K$, we make the same assumptions and follow the same optimization procedure as with Fig. \ref{fig_m_scaling}. When $K$ is increased, the system EE increases because higher throughputs can be achieved when more UEs are scheduled. However, since an increase in $K$ also results in increased circuit power consumption, we observe that the per-unit increase in EE decreases with $K$. Additionally, we observe that the system EE achieves different peak levels for different combinations of $M$ and $K$. Based on this observation, we can design traffic-adaptive antenna selection methods, where the BS can dynamically activate a subset of its antennas with changing traffic demands in the system. The BS can navigate along the antenna selection curve in Fig. \ref{fig_antenna_selection} for sustained operation at peak EE levels, even if traffic demands vary with time.

Observe that, depending on the subset selection algorithms, such as, orthogonal matching pursuit and gradient descent, antenna selection methods may add to the computational burden at the BS. As a result, EE maximization using antenna selection becomes a non-trivial task. Current literature on antenna selection for massive MIMO is mostly confined to simple single cell scenarios. Performance tradeoffs introduced by design limitations, such as CSI availability, pilot contamination, and antenna correlation, are not clearly understood.

\subsubsection {Redesign transceiver architecture} \label{transceiver_redesign}
An alternative strategy to reduce RF chain requirements at the BS is to redesign the BS transceiver architecture. In this direction, few {\em single RF chain transceivers} have been recently designed, although at the cost of some serious practical limitations. For example, the electronically steerable parasitic antenna array proposed in \cite{MM24} operates with a single RF chain but supports a limited set of modulation schemes and requires almost twice the number of antennas than in conventional transceivers. Similarly, \cite{MM25} proposes a {\em single RF chain transmitter} based on a two-port matching network, but the transceiver performance is subject to power losses in the matching network and mutual coupling in the antenna array. Consequently, although transceiver redesign offers great promise to improve EE in MM networks, current literature cannot be considered complete. Further research is required on addressing several design issues and on overcoming any implementation challenges thereof.

In the next section, we discuss EE maximization in ``hybrid MM networks" where massive MIMO operates in conjunction with other promising 5G technologies, namely, millimeter wave, heterogenous networks, and energy harvesting networks.

\section{Hybrid Massive MIMO Networks for Energy-Efficiency in a 5G Architecture}

Hybrid MM networks can achieve higher EE levels than conventional MM systems because massive MIMO benefits mutually from the 5G technologies mentioned above. As will be discussed in the next few subsections, hybrid MM networks exhibit some unique properties and are subject to some new design constraints, thus opening up multiple opportunities and challenges for the design of energy-efficient systems. 

\begin{figure*}
\vspace{-3mm}
\begin{center}
\includegraphics [scale=0.7] {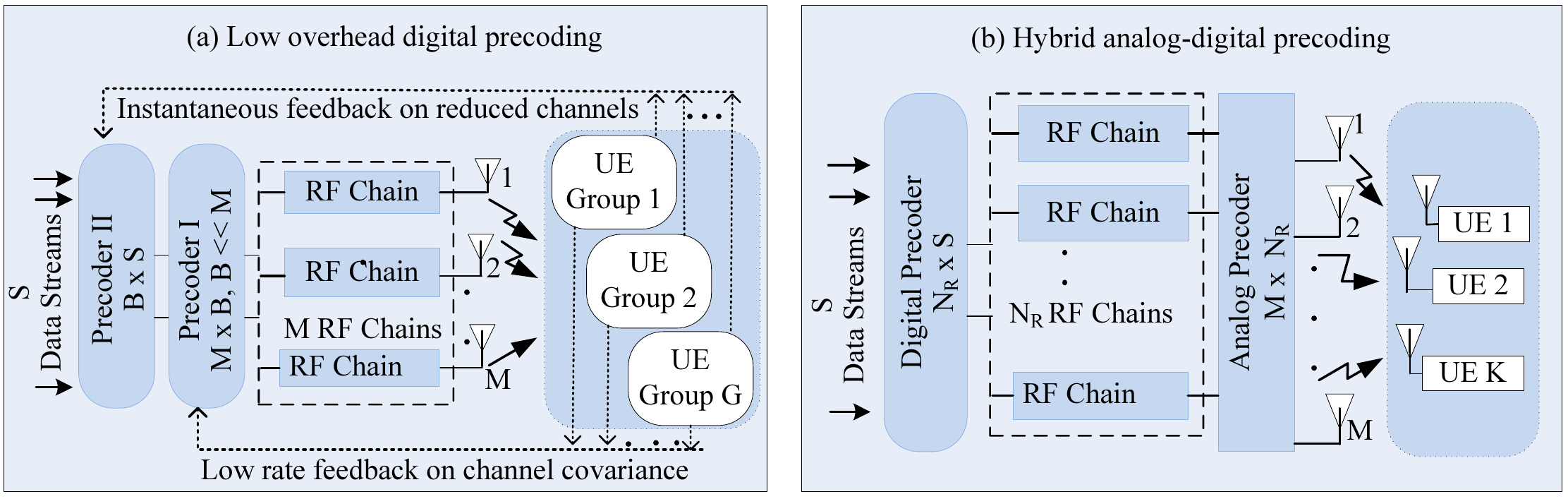}
\caption{Novel precoding techniques for mmWave massive MIMO systems.}
\label{fig_mmwave_precoding}
\end{center}
\vspace{-3mm}
\end{figure*}

\subsection{Millimeter Wave (mmWave)-Based MM Systems} \label{sec_mmwave}

The mmWave spectrum, which refers to spectrum in the 3-300 GHz band, is now being investigated for 5G operations because the current sub-3GHz bands have become overcrowded and there is a need for additional spectrum to accommodate future traffic demands. By moving to the mmWave spectrum, significant throughput gains and latency reductions can be achieved because large bandwidths of the order of multiple GHz are available $-$ bandwidths upto 7 GHz are available in the 60 GHz band. Typically, mmWave channels exhibit huge reflection and absorption losses, poor diffraction characteristics, and low channel coherence times. As a result, when compared to sub-3GHz bands, mmWave channels experience much higher channel correlation, signal attenuation, and sensitivity to blockage. mmWave channel estimation is still an active topic of research. 

\subsubsection {Benefits from co-existence} Massive MIMO implicitly offers the highly directional and adaptive transmissions required to improve signal strength and suppress interference in the blockage-sensitive environments at mmWave bands. On the other hand, mmWave makes massive MIMO realizable because (i) the small wavelengths at mmWave frequencies allow a large number of antennas to be fit into very small form factors \cite{MM4}, and (ii) the near-LOS channels in mmWave MM networks can be estimated using direction of arrival (DoA) of the incident waves at the BS, thus potentially eliminating the need for pilot reuse and the resulting pilot contamination.

\subsubsection {Opportunities for energy-efficient design} Low-complexity channel equalization techniques are sufficient for mmWave MM networks because the narrow directional beams and the near-LOS propagation eliminate much of the multipath. For the same reason, mmWave MM BSs can support low power backhaul operations, thus becoming an energy-efficient alternative to power-greedy fiber backhaul. In addition, the FDD mode of operation, which incurs large CSI overhead and is therefore impractical at the sub-3GHz bands, becomes realizable at mmWave bands because sparsity in mmWave channels can be exploited to derive low overhead precoding techniques, such as shown in Fig. \ref{fig_mmwave_precoding}(a).

The two-stage digital precoding technique in Fig. \ref{fig_mmwave_precoding}(a) exploits channel sparsity to partition UEs into different groups, each group comprising UEs with approximately the same channel covariance eigenspace, such that the covariance eigenspaces of different UE groups are near-orthogonal to each other. To understand how the CSI overhead is reduced, let us first denote $r$ as the rank of channel covariance matrix and $S$, where $(S\leq K)$, as the number of independent streams to be transmitted to the UEs. Precoder I exploits the near-orthogonality of covariance eigenspaces to reduce the channel dimensionality from $M \times K$ to $B \times S$, where $B$  ($S \leq B < r$) is an optimization parameter to regulate intergroup interference in the system. A low-rate feedback mechanism is sufficient to update precoder I because it depends only on the channel covariance, which typically varies very slowly when compared to the channel coherence time. Precoder II employs simple linear precoding techniques on the effective $B \times S$ channel so as to extract multiplexing gains within each UE group. To update precoder II, the BS should acquire instantaneous CSI of the effective $B \times S$ channel during each coherence interval. Observe that the CSI overhead will still be significantly lower than in conventional FDD systems because the overhead comes predominantly from estimating reduced dimensional channels.

Sparsity in mmWave channels can also be exploited to design hybrid analog-digital beamforming techniques which relax RF chain requirements in the system. For example, the hybrid precoding technique shown in Fig. \ref{fig_mmwave_precoding}(b) reduces the number of RF chains from $M$ to $N_R$, where $S \leq K$, $S \leq N_R \leq M$. The analog precoder applies phase-only control to extract large array gains and to reduce the channel dimensionality from $M$ x $K$ to $N_R$ x $S$. The digital precoder applies simple linear precoding techniques on the effective $N_R \times S$ channel to extract multiplexing gains. RF chain requirements are reduced because the digital precoders operate only on the effective reduced dimensional channel. 

\subsubsection{Challenges and open problems}
Despite clear evidence that multi-stage beamforming techniques, such as shown in Fig. \ref{fig_mmwave_precoding}(a), can be designed to reduce training overhead in mmWave MM systems, such techniques have only been studied to a limited degree of extent (see \cite{MM4} for an example). Tradeoffs introduced by pilot contamination are not clearly understood. Missing in the existing literature are studies which optimize the interference mitigation parameter $B$ for energy efficiency. Other open problems include optimizing user grouping, covariance tracking, and inter cell interference mitigation for energy efficiency. Similar is the situation with hybrid analog-digital beamforming techniques which relax RF chain requirements at the BS. These techniques are invaluable for mmWave operations because mixed signal components in the RF chain, particularly the high resolution ADCs, consume unacceptably large amounts of power when operated at large bandwidths. Notice that the analog precoding phase introduces several new constraints in the transceiver design, such as limited precision for phase control, limited number of phase shifts, and limited analog to digital converter resolution. Existing literature does not discuss the EE tradeoffs introduced by these new constraints, leaving huge scope for further research. 

Another major bottleneck in the realization of energy-efficient mmWave MM systems is the hardware design. Silicon-based CMOS technologies provide a simple and cost-effective means to integrate several mmWave antennas with necessary analog and digital circuitry onto a single package. However, the high frequency and large bandwidth operations in the mmWave regime impose several constraints on the design of transceiver components. For example, high substrate absorption losses and high noise power levels become roadblocks to the design and integration of highly directional antennas into CMOS packages. In addition, improper isolation between active on-chip components can result in mutual coupling, self-jamming, and signal distortion. Transceivers which address these design complications have not been fabricated till date.

\subsection{MM-Based Heterogenous Networks}
Dense heterogenous networks (HetNets), where spectrum utilization is maximized by decreasing the cell size and increasing the number of small cells (SCs) per unit area, offer a promising approach to satisfy the traffic demands expected in 5G. In terms of energy efficiency, HetNets are a superior alternative to massive MIMO because (i) the power consumption per small cell access point (SCA) is generally low, (ii) SCAs can be opportunistically turned on/off depending on traffic demand, and (iii) high throughput gains can be achieved by intelligently offloading traffic between outdoor and indoor SCs. Moreover, when $M$ SCs are deployed per unit area and $\gamma$ is the path loss exponent, $O(M^{\frac {\gamma}{2}})$ array gains can be achieved because the average BS-UE distance is reduced by $M^{\frac {1}{2}}$. These array gains are larger than the $O(M)$ gains offered by massive MIMO because $\gamma > 2$ for most propagation conditions. 

\subsubsection{Benefits from co-existence} Due to smaller coverage areas, SCs fail to ensure seamless connectivity and quality of service (QoS) to UEs which are highly mobile. This limitation can be overcome by designing a two-tier MM HetNet, wherein a macro cell tier formed by the MM BSs is overlaid with an SC tier formed by small cells, such as pico cells and femto cells. The macro cell tier ensures uniform service coverage and supports highly mobile UEs, while the small cell tier caters to the local indoor and outdoor capacity requirements. Clearly, such an architecture can simultaneously extract the $O(M^{\frac {\gamma}{2}})$ array gains offered by HetNets and the $O(K)$ multiplexing gains offered by massive MIMO. In addition, since the macro tier hosts a large number of antennas, few antennas can be reserved for low power wireless backhaul to the SC tier. Interference coordination in MM HetNets can be analyzed by using simple tools from random matrix theory. This is highly beneficial because tools from stochastic geometry, which are used to study interference coordination in single antenna HetNets, cannot be easily applied to MM HetNets because beamforming introduces cross-tier statistical dependencies.

\begin{figure}
\begin{center}
\includegraphics [scale=0.5] {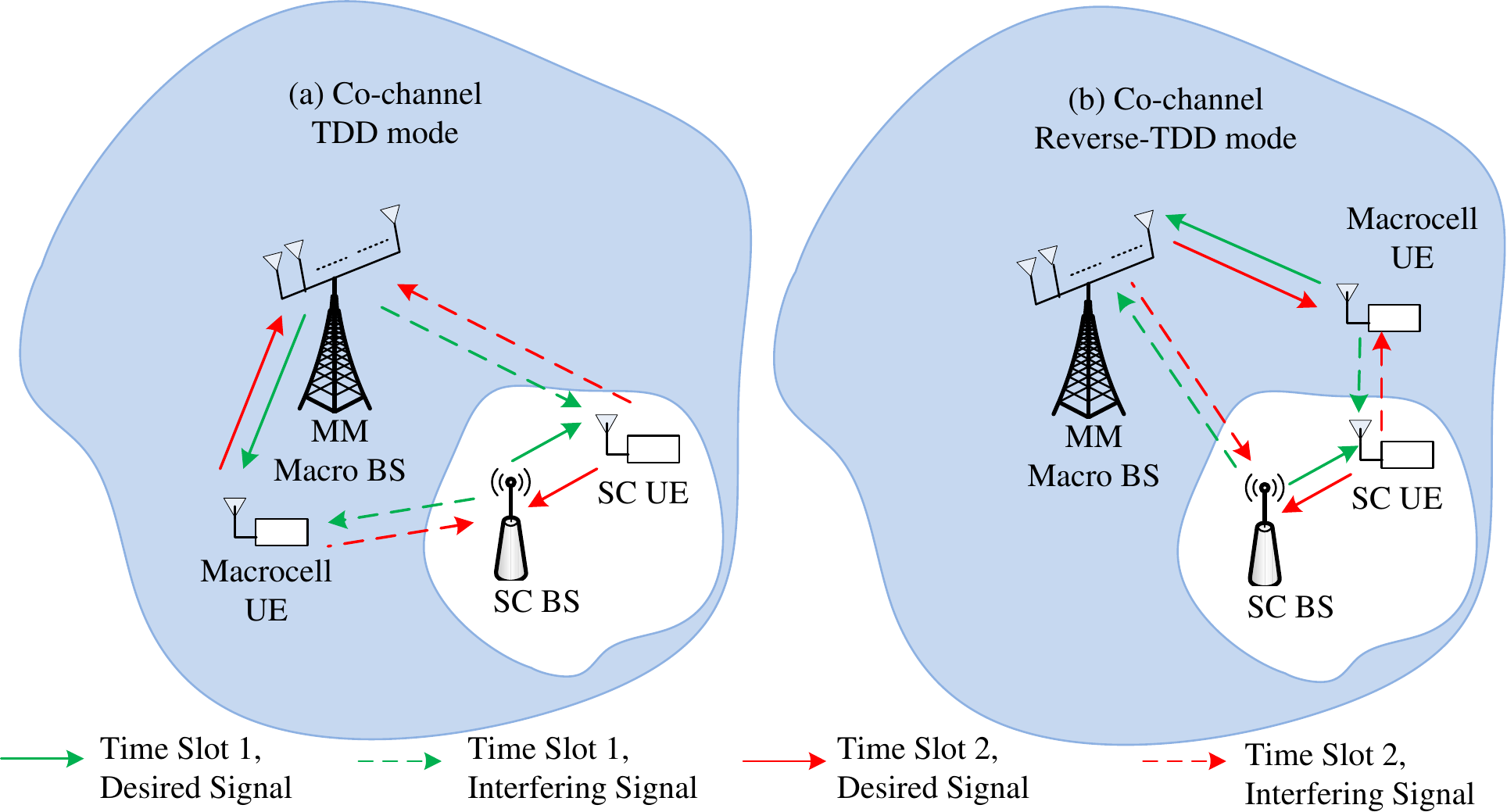}
\caption{Co-channel TDD and co-channel reverse TDD deployment modes for massive MIMO HetNets.}
\label{fig_MM_Hetnets}
\end{center}
\end{figure}

\subsubsection {Opportunities for energy-efficient design}
Energy efficiency in MM HetNets can be improved by combining the EE maximization techniques discussed in Section \ref{ee_design_methods} with few EE maximization techniques for HetNets, such as, BS sleeping, cell zooming, cell association, and coordinated multi-point transmission (CoMP). Several other energy-efficient techniques can be designed by jointly exploiting the properties of MM and HetNet technologies. For example, MM HetNets can use low-complexity multi-flow beamforming techniques to jointly coordinate interference among UEs in both macro and SC tiers. Such techniques are known to drastically reduce hardware requirements at the MM BSs $-$ \cite{MM66} shows that the number of MM BS antennas can be reduced by more than 50\% if a few single antenna SCs are overlaid on the MM cell.

In addition, co-channel deployment modes, where the available spectrum is fully utilized in both macro and SC tiers, can be attempted. Example scenarios are illustrated in Fig. \ref{fig_MM_Hetnets} as the co-channel TDD (co-TDD) and the co-channel reverse TDD (co-RTDD) modes. In the co-TDD mode, the macro and SC tiers are time-synchronized to simultaneously transmit in the UL or the DL. In the co-RTDD mode, the order of UL and DL transmissions are reversed in one of the tiers, i.e., macro tier operates in the DL when SC tier operates in the UL and vice versa. Since the entire spectrum is utilized in both the tiers, simultaneous and uncoordinated transmissions can introduce significant inter-tier and intra-tier interference. Fortunately, the BSs in macro and SC tiers can not only estimate the channels to their intended UEs, but also the covariance of interfering signals. As a result, channel reciprocity can be exploited to design precoding vectors which sacrifice certain degrees of freedom (DoFs) on the DL so as to blank out the strongest interference subspace. When such {\em spatial blanking} techniques are used and the number of sacrificed DoFs are optimized, significant throughput gains can be achieved in the SC tier at the cost of a negligible throughput loss in the macro tier \cite{MM61}.

The co-TDD and co-RTDD modes exhibit conflicting properties, leading to some interesting tradeoffs during the design of energy-efficient MM HetNets. For example, the quality of interference estimation and the ability to reject interference can be considerably different because the interfering signals are radically different. Co-RTDD renders higher interference estimation accuracy than co-TDD because the interferer channels are quasi-static in co-RTDD, due to fixed locations of the MM and SC BSs, but are dynamically varying in co-TDD, due to moving UEs. Consequently, when in the macro-tier UL, co-RTDD can attempt {\em spatial blanking} to achieve higher throughput gains than co-TDD. On the other hand, when in the macro-tier DL, co-RTDD offers lower throughput gains than co-TDD because co-RTDD renders lower interference rejection. This is in turn because the SCs have much fewer antennas than the MM BSs and hence, sacrificing DoFs at the SCs may not reduce the cross-tier interference significantly. EE gains offered by co-TDD and co-RTDD are also subject to UE density in the network. For example, analytical studies \cite{MM62} show that higher throughputs are achieved if more SCs operate in co-RTDD for sparse networks and in co-TDD for dense networks.

\subsubsection{Challenges and open problems} Several challenges continue to roadblock the design of energy-efficient MM HetNets. For example, most studies on {\em spatial blanking} attempt channel covariance estimation and precoding based on a wide sense stationarity assumption on the channel process. Such an assumption is valid only locally and is susceptible to mobility in the system. Therefore, novel channel tracking algorithms should be developed to adaptively learn and update the estimated interference subspace according to the non-stationary time-varying effects in the system. Also, most studies on {\em spatial blanking} focus on simplistic UE distribution scenarios with either isolated UEs or hotspots. In contrast, realistic HetNets would experience asymmetric traffic loads coming from a combination of hotspots and isolated UEs. Therefore, advanced low complexity interference coordination strategies should be designed to allow efficient spatial resource sharing between hotspots and isolated UEs. Additionally, as discussed earlier, there is no clear winner among co-TDD and co-RTDD. This calls for the design of innovative co-channel deployment modes, which can simultaneously reap the benefits and overcome the limitations of co-TDD and co-RTDD. Appropriate pilot assignment methods should be developed to contain pilot contamination, which can be particularly severe in co-channel deployments. Load balancing in MM HetNets is another largely unexplored subject. Resource efficient inter-tier offloading techniques based on load-adaptive cell zooming, dynamic antenna activation in the macro tier, and mobility-aware handover policies, should be designed under practical constraints such as limited backhaul and load asymmetries. 

\subsection {Energy Harvesting (EH)-Based MM Networks}
\subsubsection{ Benefits from co-existence} Powering the BSs in MM networks with energy harvested from renewable resources, such as solar, wind, and thermal, can result in reduced carbon footprint and increased network lifetime. The UEs in an MM network can also benefit from EH capabilities because they are usually powered by limited-capacity batteries. Unlike at the BSs, EH rates at the UEs should be controllable because battery drains can potentially lead to loss of network connectivity. To achieve this, the BSs can transmit dedicated RF signals on the DL and perform energy beamforming so as to provide uninterrupted wireless energy transfer (WET) to the UEs. Alternatively, the BSs can attempt simultaneous wireless information and power transfer (SWIPT), where the DL RF signals are used to simultaneously transport both energy and information to the UEs. MM is particularly suitable for such RF EH applications because the large array gains offered by MM can increase the energy transfer efficiency of RF signals. 

Incorporating EH capabilities also introduces several new design constraints into the system. For example, unlike in grid-powered networks, the energy harvested from most renewable resources fluctuates randomly, although over a smaller range of magnitude and a larger timescale than the communication channel amplitudes. Signal interference may not be desirable from a throughput perspective, but is desirable from an RF EH perspective. Also, power sensitivity levels are radically different for RF energy harvesters (about $-10$ dBm) and information receivers (about $-60$ dBm). In addition, EH networks are subject to a new causality constraint: energy consumed until a given time cannot exceed the energy harvested until then. These new constraints in EH networks enforce the transmission policies to depend not only on channel fading but also on factors such as energy arrival and data backlog.

\begin{figure}[h]
\begin{center}
\includegraphics [scale=0.6] {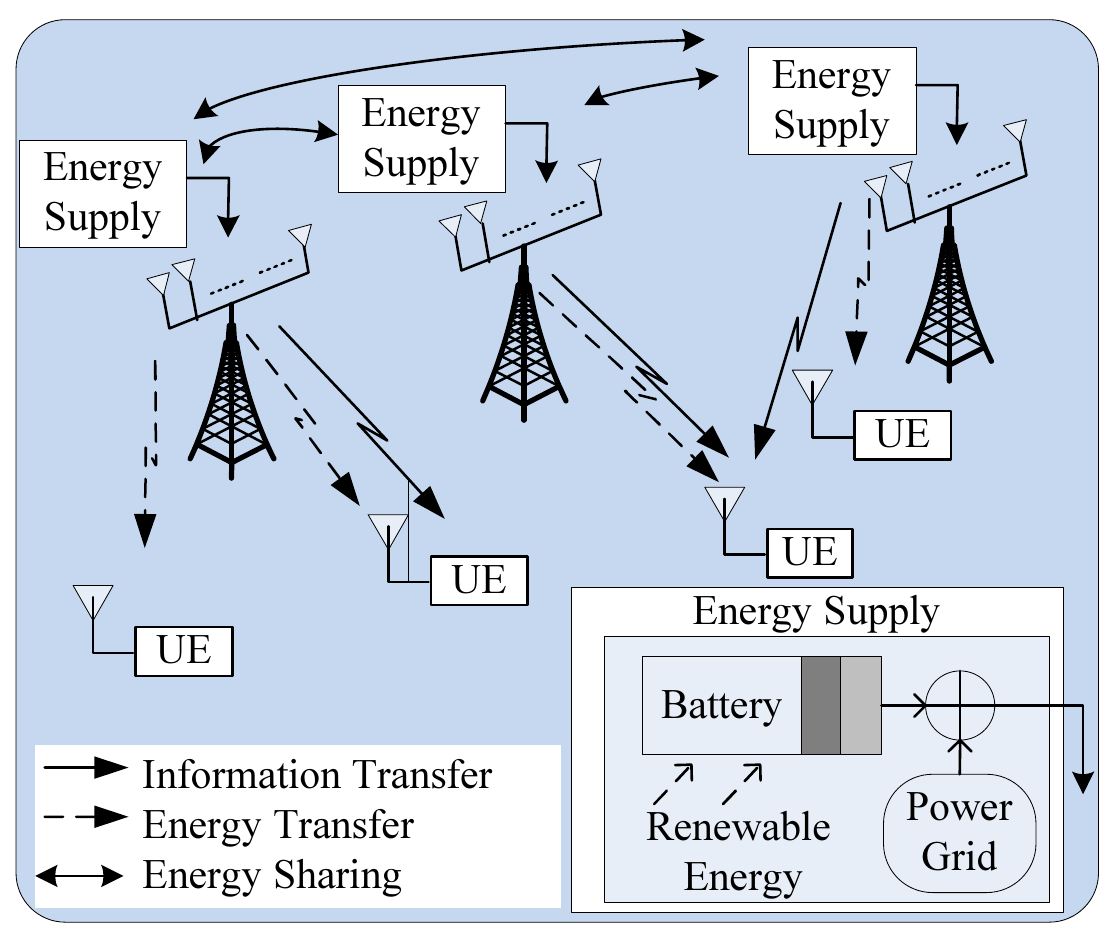}
\caption{Massive MIMO with energy harvesting capability.}
\label{fig_MM_EH_nets}
\end{center}
\end{figure}

\subsubsection {Opportunities for energy-efficient design} Energy harvesting is a natural option to improve EE in MM networks because it opens up several new opportunities for minimizing power consumption from non-renewable resources. Consider, for example, the EH MM network illustrated in Fig. \ref{fig_MM_EH_nets}, where the MM BSs are powered by energy supplies which can harvest and store renewable energy in a battery. Since the energy harvested from renewable sources is generally sporadic, the energy supply at the BS can optionally draw power from the grid so as to ensure network reliability. In addition, since the EH rates may be different for different BSs, the MM network can employ an energy sharing architecture, which allows the energy harvested locally at each BS to be shared across the network. Thereby, the BSs can implement transmission policies which optimally utilize the harvested energy. The BSs can also attempt WET or SWIPT to enable RF energy harvesting at the UEs. Significant EE gains are achieved because the UEs can use the harvested energy to recharge batteries or to power UL transmissions.

\subsubsection{Challenges and open Problems}
Despite the well-known benefits of using a large number of antennas at the BS, limited literature is currently available on EH MM networks. In addition, most transmission policies available for EH MIMO networks cannot be generalized to the MM regime because these policies either assume the circuit power consumption, i.e., $P_C$, to be zero or assume $P_C$ as a constant term which is independent of $(M, K)$. As discussed earlier in Section \ref{EE_define}, such assumptions on $P_C$ are not valid in the MM regime. Very few studies, such as \cite{MM100}, use a realistic model for $P_C$ but make other unrealistic assumptions, such as, perfect CSI and full knowledge of future energy arrival. This leaves huge scope for future work, particularly on the EE tradeoffs introduced by constraints, such as, battery imperfections, delay-sensitive traffic, and lossy energy sharing architectures. 

RF energy harvesting capabilities have also not been thoroughly investigated in MM networks. Few studies, such as \cite{MM108}, study energy beamforming for WET in MM networks under practical constraints such as imperfect CSI, delay-sensitive traffic, and frequency-selectivity of channels. However, tradeoffs introduced by several other realistic constraints, such as, finite battery capacity, energy leakages, and transceiver imperfections, are yet to be fully understood. Few other studies \cite{MM112} discuss SWIPT-enabled MM networks, but several important concerns are yet to be addressed. For example, information leakage concerns, which arise when DL signals are amplified to improve RF EH rates at the UEs, have not been addressed so far. Energy-aware medium access control (MAC) protocols, which optimize time resource allocation between channel access and EH, are yet to be designed.

Several interesting research directions should be pursued in the future to fully realize the EE gains offered by EH MM networks. Cross-layer design techniques should be developed to allow the BSs to jointly schedule energy and data transmissions based on channel conditions. Traffic-adaptive transceiver activation methods should be designed to efficiently utilize the energy harvested at the BS. Another important research direction is to optimize pilot power and pilot symbol placement under EH constraints. Reliability concerns in EH MM networks should be addressed through the use of hybrid power supplies, cooperative relays, and energy cooperation techniques. Another major research direction is to study optimal energy management policies in the presence of uncertainties in the battery state information $-$ circuit components are known to introduce uncertainty errors as high as 30 \%. In this context, partially-observed Markov decision process frameworks may be utilized to arrive at intelligent trade-offs between the accuracy of battery state information and the amount of energy spent in acquiring it. Lastly, EH devices and communication protocols should be standardized so as to simplify network planning and management.

\section {Conclusion}
Massive multiple-input multiple-output (MIMO) is a promising technology for sustainable evolution towards 5G because it offers multiple orders of spectral and energy efficiency (EE) gains over current LTE technologies. This article has explored several opportunities to boost the EE gains offered by massive MIMO (MM). Standard techniques for EE enhancement in MM systems, such as scaling the number of BS antennas, implementing low-complexity operations at the BS, minimizing power amplifier losses, and minimizing RF chain requirements, have been briefly discussed to identify few open research challenges. The article has also discussed ``hybrid MM systems", where MM operates alongside other 5G technologies, such as millimetre wave, heterogenous networks, and energy harvesting networks. The mutual benefits and the performance trade-offs introduced by the coexistence of these 5G technologies have been discussed. Several new opportunities and open research challenges have been identified for the design of energy-efficient hybrid MM systems. Our discussions show clear indications that hybrid MM systems have enormous potential to meet the energy-efficiency demands expected in 5G cellular networks.

\bibliographystyle{IEEEtran}

\ifCLASSOPTIONcaptionsoff
  \newpage
\fi

\end{document}